%% file: masterfile.tex
\documentclass[10pt,twoside,BCOR7mm,DIV15,headinclude,footexclude,
               cleardoubleempty,idxtotoc]
{scrartcl}

\usepackage{natbib}
\usepackage[font=small,labelfont=bf]{caption}
\usepackage[english]{babel}
\usepackage{graphicx}
\usepackage{hyperref}
\usepackage{scrpage2}
\usepackage{ifthen}
\usepackage{booktabs}
\usepackage{amsmath}
\usepackage{amssymb}
\usepackage{multicol}
\usepackage{float}
\usepackage{hyperref}

\hypersetup{breaklinks=true
,colorlinks=true,linkcolor=black,urlcolor=blue
,citecolor=black}

\addto\captionsenglish{%
}

\makeatletter
\renewcommand{\@biblabel}[1]{}
\renewcommand{\@cite}[2]{%
{#1\ifthenelse{\boolean{@tempswa}}{,#2}{}}}
\makeatother
\ofoot{\thepage}
\ifoot{}

\setcapindent{0em}
\setheadsepline{1pt}

\setkomafont{pagehead}{\normalfont\sffamily}
\setkomafont{pagenumber}{\normalfont\rmfamily}

\makeatletter
\newcommand{\listofcontributions}{\@starttoc{con}}

\newcommand{\l@contribution} {\@dottedtocline{1}{1.5em}{2.3em}}
\makeatother

\newenvironment{contribution}{
\setcounter{section}{0}
\setcounter{figure}{0}
\setcounter{table}{0}
}{
\newpage
\lehead{}
\rohead{}
}

\begin{document}

\setlength{\baselineskip}{2.5ex}

\begin{contribution}
\include{myarticle}
\end{contribution}


\end{document}

%% file: myarticle.tex

\lehead{P.M.\ Williams \& K.A.\ van der Hucht}

\rohead{The colliding-wind WC9+OB system WR 65}

\begin{center}
{\LARGE \bf The colliding-wind WC9+OB system WR 65 and dust formation by WR stars}\\
\medskip

{\it\bf P.M. Williams$^1$ \& K.A.\ van der Hucht$^2$}\\

{\it $^1$University of Edinburgh, Royal Observatory, Edinburgh, United Kingdom}\\
{\it $^2$SRON Utrecht and University of Amsterdam, The Netherlands}

\begin{abstract}
Observations of the WC9+OB system WR\,65 in the infrared show variations of its 
dust emission consistent with a period near 4.8~yr, suggesting formation in a 
colliding-wind binary (CWB) having an elliptical orbit. If we adopt the IR maximum 
as zero phase, the times of X-ray maximum count and minimum extinction to the hard 
component measured by Oskinova \& Hamann fall at phases 0.4--0.5, when the 
separation of the WC9 and OB stars is greatest. We consider WR\,65 in the context 
of other WC8--9\,+\,OB stars showing dust emission.

\end{abstract}
\end{center}

\begin{multicols}{2}

\section{Identification of WR 65 as a colliding-wind binary}

From 1.2--9.7-$\mu$m infrared (IR) photometry, \citet{WHT} found that 
WR\,65 (LSS 3319), like most of the then-known WC9 stars, had a 
circumstellar dust cloud which, in this case, absorbed and re-emitted 
about 3\% of its stellar flux in the IR. 

\begin{figure}[H]        
\begin{center}
\includegraphics[width=\columnwidth]{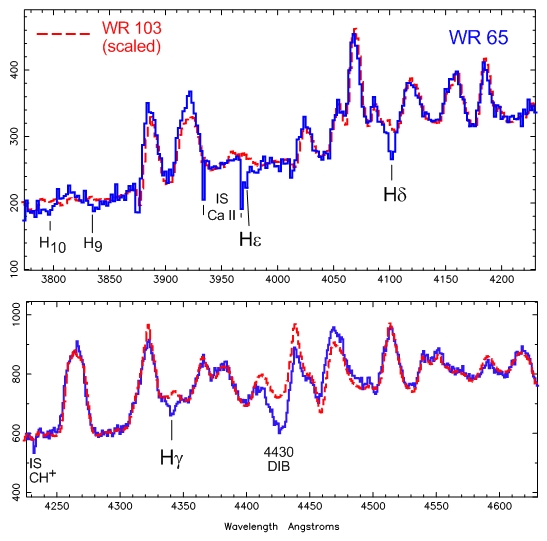}
\caption{Spectra of WR\,65 and WR\,103 observed with the ESO EMMI on the 
NTT. The interstellar lines in WR\,65 are stronger on account of its higher 
reddening ($A_v \simeq 7.6$).}
\label{WilliamsF1}
\end{center}
\end{figure}

Spurred by the episodic formation of dust by the WC7+O5 binary WR\,140 \citep{W140} 
at periastron when its components reached a critical separation, we suggested 
\citep{WASP22} that the persistent WC8--9 dust makers might also be binaries,  
but in circular orbits having stellar separations conducive to dust formation --- 
as beautifully confirmed by the rotating dust `pinwheel' made by the prototypical 
WC9 dust maker WR\,104 \citep{Tuthill104} --- and we began a spectroscopic 
survey of WC9 stars in the blue using the SAAO 1.9-m telescope and the ESO NTT 
to search for absorption lines attributable to OB companions. 
Amongst the WC9 stars found \citep{WLiege05} to show Balmer absorption lines 
in their spectra was WR\,65 (Fig.\,\ref{WilliamsF1}), making it a candidate SB2. 
From dilution of the WC9 emission-line spectrum, \citet{VIIthCat} suggested 
that the OB star was 0.4~mag. brighter than the WC9 star.
 
Evidence that WR\,65 might be a colliding-wind binary came from the variable X-ray 
emission observed by \citet{Oskinova65}. They fitted a $2T2N_H$ plasma model, 
and found that the column density to the `hard' component varied with epoch, 
consistent with the movement of an embedded colliding-wind source in 
the WC9 wind.

\begin{figure}[H]  
\begin{center}
\includegraphics[width=\columnwidth]{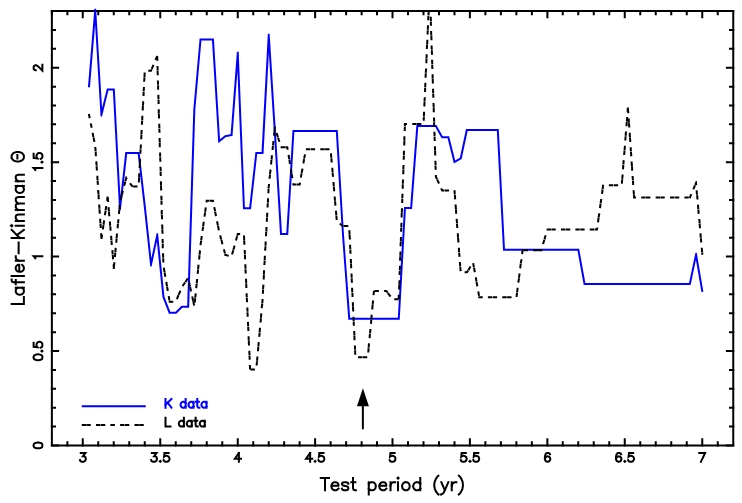}
\caption{Lafler-Kinman period search on $K$ and $L$-band photometry 
with adopted 4.8-yr period marked ($\uparrow$).}
\label{WilliamsF2}
\end{center}
\end{figure}

We therefore examined the IR photometric history of WR\,65. Besides the 
1983 observations in \citet{WHT} and those by \citet{PEG} in 1982, eight 
more sets of $JHKL$ photometry were observed at ESO during 1990--93 as 
part of another programme. Four of these observations were taken on 
successive nights in 1991 and the last of them shows brightening 
(0.09-mag.\ in $JHK$, less in $L$ and $M$), suggesting a mini-outburst 
like those observed from WR\,137 \citep{W137II}. Latterly, $iJK_s$ were 
observed in the DENIS survey \citep{DENIS} in 1998 and $JHK_s$ in the 
2MASS survey \citep{2MASS} in 1999. These $J$ and $K_s$ magnitudes are 
significantly brighter than any of the earlier $J$ and $K$, suggesting 
occurrence of an IR maximum in 1998--99. For comparison with the $L$ 
data, we determined [3.6] magnitudes from the pipeline-processed 
{\em Spitzer} IRAC frames observed in 2004 and 2012. We searched for 
a period in these rather sparse data following \citet{LaflerKinman}. 
As the dates of the DENIS and 2MASS $K_s$ band observations were different 
from those of the [3.6] magnitudes, there is a small difference in the 
cadence of the $K/K_s$ and $L$/[3.6] datasets and we sought matching 
minima in the L--K statistic $\Theta$ from both of them. 
This is shown in Fig.\,\ref{WilliamsF2}, 
from which we adopt a tentative period of 4.8$\pm$0.2 yr.

\begin{figure}[H]  
\begin{center}
\includegraphics[width=\columnwidth]{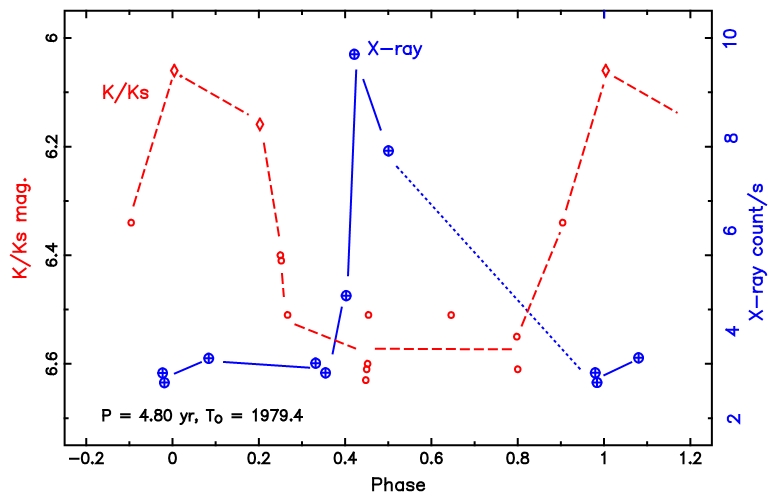}
\caption{IR $K$ ($\circ$) and $K_s$ ($\diamond$) photometry together 
with X-ray count rates ($\oplus$) from \citet{Oskinova65}, all phased 
to a period of 4.8 yr and taking 1979.4 as zero. (The straight 
lines joining the observations are drawn just to guide the eye.)}
\label{WilliamsF3}
\end{center}
\end{figure}

The phased IR ($K$) magnitudes and X-ray count rates are plotted in 
Fig.\,\ref{WilliamsF3} using our IR period of 4.8~yr and arbitrarily 
adopting $T_0$ = 1979.4, so as to place the brightest IR flux, the 
DENIS observation, at phase zero. 
The IR maximum is broad but poorly defined; more IR observations 
are needed to define the IR light curve and period. At minimum, the 
IR spectral energy distribution (SED) still shows dust emission. 
The X-ray emission remains faint during IR maximum, when the 
wind-collision source would have been most deeply buried in the WC9 wind. 
The rise to X-ray maximum correponding to minimum column density to the 
`hard' component occurs as the system approaches apastron, but the 
steepness of the rise suggests that we are observing it through a hole 
in the WC9 wind formed by the O star wind sweeping through our 
line-of-sight. Further X-ray observations are needed to define the 
light curve and map the wind-collision region (WCR). 
Above all, we need an orbit to allow a comprehensive picture to be developed.
The period near 4.8~yr makes WR\,65 a tractable source for a study 
on a reasonable time-scale; it is the only WC9 system known to have 
periodically variable dust emission (cf. Table \ref{WilliamsTP}).

Colliding stellar winds can also produce non-thermal radio emission, 
observable if the WCR lies sufficiently far out in the stellar wind, 
especially in wide, long period systems. The observable emission may 
vary during the orbit \citep{DWnT} in systems of intermediate period. 
Our suggested 4.8-yr period lies in that range. The marginal 3.5-cm 
detection (0.37$\pm$0.14~mJy) by \citet{LeithererRadio} is consistent 
with the free-free emission from the wind of WR\,65, 0.35~mJy, which 
we estimated by scaling the observed flux from $\gamma$~Vel to the 
distance (3.3~kpc) of WR\,65. The date of the observation corresponds 
to phase 0.31 on our suggested ephemeris, when the low X-ray flux 
implies that the wind extinction to the WCR and any non-thermal radio 
source is still high. We suggest that re-observation at a more 
favourable phase may well reveal non-thermal radio emission from WR\,65.

\begin{table}[H] 
\begin{center} 
\captionabove{Wolf-Rayet stars showing periodic dust emission. The IR 
photometric periods of WR\,140 and WR\,137 are supported by RV periods 
\citep{Fahed140,Orbit137} and that of WR\,98a by its pinwheel 
rotation \citep{Monnier98a}.} 
\label{WilliamsTP}
\begin{tabular}{cccc}
\toprule
   WCE     &     WC7     &      WC8     &     WC9      \\ 
   (P/yr)  &   (P/yr)    &     (P/yr)   &   (P/yr)     \\
\midrule
 HD 36402  &    WR 140   &     WR 98a   & {\bf WR 65}  \\
 (4.7)$^a$ & (7.94)$^b$  &  (1.54)$^c$  & ($\sim$ 4.8) \\ 
 WR 19     &   WR 137    &     WR 48a   &              \\
(10.1)$^d$ & (13.05)$^e$ & ($\sim$ 32)$^f$ &           \\
\bottomrule
\end{tabular}
\end{center}
Refs: $^a$ \citet{W36402}; $^b$ \citet{W140}; $^c$ \citet{WIAU212};
$^d$ \citet{Veen19}; $^e$ \citet{W137II}; $^f$ \citet{W48a}
\end{table}

\section{Relation to other persistent dust-making WC8--9 stars.}

\begin{figure}[H]  
\begin{center}
\includegraphics[width=\columnwidth]{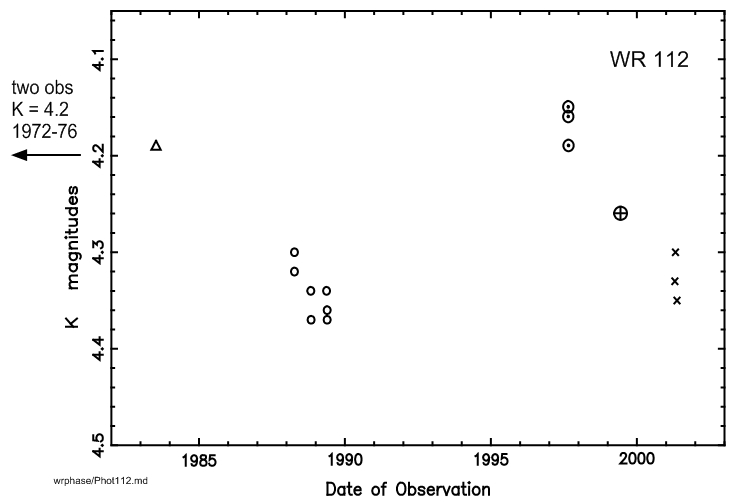}
\caption{Photometry of WR\,112 from \citet{WHT} ($\triangle$), 
new observations at ESO ($\circ$) and with the Carlos S\'anchez IR 
Telescope ($\odot$), from the 2MASS Catalogue ($\oplus$) and by 
\citet{Kimeswenger} ($\times$). We do not have dates for the earliest 
observations by \citet{MS76} and \citet{AHL77}.}
\label{WilliamsF4}
\end{center}
\end{figure}

From their `pinwheel' images of WR\,104, \citet{Tuthill104b} showed 
that the dust was made by stars moving in a circular orbit and that 
the IR flux, and hence dust formation rate, did not vary with 
orbital phase. 
Also, the long-term (1982--1997) photometric history of WR\,104 from 
14 observations show a dispersion of only $\Delta K = 0.04$ mag. 
We have compiled long-term IR photometric histories of 12 other 
WC8--9 dust-makers. The only one to show systematic variability 
is WR\,112 (CRL 2104), which may have a period near 12.3~yr 
(Fig.\,\ref{WilliamsF4}). This period is consistent with its dust 
`pinwheel' and a plausible expansion velocity \citep{Marchenko112} 
with a revised distance $\sim$ 2~kpc \citep{Marchenko48a112}. 
The low orbital eccentricity inferred from its dust `pinwheel' is 
consistent with the low amplitude of its IR variations compared 
with those ($\Delta K > 1$) of the high eccentricity binaries 
WR\,140 \citep{Fahed140} and WR\,19 \citep{W19}.

The long-term photometry of another `heavy' dust maker, WR\,118, shows 
a dispersion $\Delta K = 0.11$ mag. but with no apparent periodicity. 
Two other dust-making WC9 stars found to have Balmer absorption lines 
\citep{WLiege05}, WR\,59 and WR\,69, have smaller dispersions in their 
photometric histories and we suggest that they are binaries in (near) 
circular orbits. The same might apply to other WC9 stars showing steady 
dust emission, which might be undetected binaries, especially if the 
WC9 stars are more luminous than currently thought, making it harder to 
detect composite spectra. There is a strong 
case for RV studies of these objects to search for orbital motion.

\bibliographystyle{aa} 
\bibliography{myarticle}

\end{multicols}